\documentclass[pra,aps,epsf,nofootinbib,
notitlepage,showpacs,10pt, 
]{revtex4-1}

\usepackage{color}
\usepackage[dvipsnames]{xcolor}
\usepackage{float}
\usepackage{graphicx}
\usepackage{subcaption}
\usepackage{amsmath}
\usepackage{amsthm}
\usepackage{mathtools}

\theoremstyle{definition}
\newtheorem{thm}{Theorem}[section]

\usepackage{bm}

\newcommand{\bra}[1]{{\left\langle #1 \right|}}
\newcommand{\ket}[1]{{\left| #1 \right\rangle}}
\newcommand{\ketbra}[2]{{\left| #1 \middle\rangle \middle \langle #2 \right|}}

\newtheorem{property}[thm]{Property}

  \newcommand{\diagdots}[3][-25]{%
  \rotatebox{#1}{\makebox[0pt]{\makebox[#2]{\xleaders\hbox{$\cdot$\hskip#3}\hfill\kern0pt}}}%
}

\theoremstyle{definition}

\usepackage{graphicx} 
\usepackage{tikz} 
\usetikzlibrary{matrix,positioning}
\usepackage{amsfonts}
\usepackage{amssymb}
\usepackage{bbold}

\begin{document}

\markboth{Rebekah Herrman}
{Relating ma-QAOA and dynamic CTQWs}

\title{Relating the multi-angle quantum approximate optimization algorithm and continuous-time quantum walks on dynamic graphs}
	
\author{Rebekah Herrman}
\email{rherrma2@utk.edu}
\affiliation{
	Department of Industrial and Systems Engineering\\ University of Tennessee at Knoxville\\Knoxville, TN  37996 USA}

\begin{abstract}

In this work, we show that ma-QAOA is equivalent to a restriction of continuous-time quantum walks on dynamic graphs. We then show it is universal for computation by finding the appropriate $B$ and $C$ operators and angles that implement the universal gate set consisting of the Hadamard, $\pi/8$ and Controlled-Not gates in the ma-QAOA framework. This result begins to bridge the gap between the continuous-time quantum walk model and gate model of quantum computation.

\end{abstract}
\maketitle 

\section{Introduction}

In classical computing, combinatorial optimization (CO) problems are defined by bits and constraints on the bits, called clauses. The goal of these problems is to maximize or minimize some objective, 
\begin{equation*}
    C = \sum_a C_a,
\end{equation*}
\noindent where $C_a$ refers to the $a^{th}$ clause. The quantum approximate optimization algorithm (QAOA) is a well-studied algorithm that approximately solves CO problems \cite{farhi2014quantum}. The algorithm requires that $C$ is encoded into a unitary operator
\begin{equation*}
U(\gamma, C) =  e^{-iC\gamma},
\end{equation*}
\noindent where $\gamma \in \mathbb{R}$ is a real valued parameter, often called an angle. $U(\gamma, C)$ and a mixing unitary, 
\begin{equation*}
U(\beta, B)= e^{-iB\beta}
\end{equation*}
\noindent where $\beta \in \mathbb{R}$, are applied to an initial state which is the equal superposition over the computational basis states 
\begin{equation*}
\ket{s}=\frac{1}{\sqrt{2^n}} \sum_{z} \ket{z}.
\end{equation*}
\noindent Typically $B$ is a sum of Pauli-x operators acting on each qubit, however, other mixers have been considered \cite{rieffel2020xy, bartschi2020grover}. The QAOA ansatz applied $p$ times to $\ket{s}$ is denoted $p$-QAOA. The result of $p$-QAOA is
\begin{equation*}
\ket{\gamma,\beta}=U(\beta_p, B)U(\gamma_p,C)  \ldots U(\beta_1, B)U(\gamma_1,C)\ket{s},
\end{equation*}

\noindent where the subscript $i$ denotes the angle chosen for iteration $i$ of the algorithm. The classical parameters $\gamma$ and $\beta$ are chosen to maximize $\bra{\gamma, \beta}C\ket{\gamma,\beta}$. While QAOA is typically thought of as an algorithm that solves CO problems, it has been shown to be a universal model of quantum computation \cite{lloyd2018quantum, morales2020universality}. 
Recently, multi-angle QAOA (ma-QAOA) was introduced as a generalization of QAOA that allows for additional classical parameters \cite{herrman2022multi}. In that same paper, it was shown that ma-QAOA always performs at least as well as QAOA for optimization and can strictly outperform it in some cases. 
 
 Continuous-time quantum walks on dynamic graphs (dynamic CTQWs) were introduced in \cite{herrman2019continuous} and proven to be universal for quantum computation. In this paper, we show that ma-QAOA is equivalent to a restriction of dynamic CTQWs in Section~\ref{sec:background}. We then find the appropriate operators $B$ and $C$, and angles $\beta$ and $\gamma$ that implement the universal gate set of the Hadamard ($H$), $\pi/8$ ($T$), and Controlled-Not ($CX$) gates in Section~\ref{sec:universalset}. These gates have been determined in the dynamic CTQW framework, so the graphs and times used in that framework define the operators and angles used in the ma-QAOA framework. In Section~\ref{sec:example}, we work through the implementation of the $H$, $T$ and $CX$ gates and then close with a discussion in Section~\ref{sec:discussion}.
 

\section{Background}\label{sec:background}
In this section, we present relevant background information for ma-QAOA and dynamic CTQWs. We then discuss how ma-QAOA can be viewed as a restriction of dynamic CTQWs. 

\subsection{ma-QAOA}\label{sec:maqaoa}
Multi-angle QAOA is a generalization of QAOA in which additional classical parameter input are allowed. As with QAOA, in ma-QAOA two operators $U(\vec{\gamma_\ell},C)$ and $U(\vec{\beta_\ell}, B)$ are applied in succession to the state $\ket{s}$ which is an equal superposition over the computational basis. $U( \vec{\gamma_\ell}, C)$ and $U(\vec{\beta_\ell}, B)$ are defined as

\begin{equation*}
U(\vec{\gamma_\ell},C) =  e^{-i \sum_{a}C_a\gamma_{\ell,a} } = \prod_{a}e^{-i\gamma_{\ell,a} C_a }
\end{equation*}
\noindent and  
\begin{equation*}
U(\vec{\beta_\ell}, B) = e^{-i \sum_{v \in V(G)}B_{v}\beta_{\ell,v} } = \prod_{v \in V(G)}e^{-i\beta_{\ell,v} B_v }.
\end{equation*}
\noindent Here, $\vec{\gamma_\ell} = (\gamma_{\ell,a_1}, \gamma_{\ell,a_2},   \ldots , \gamma_{\ell, a_m} ) \in \mathbb{R}^{m}$ is the vector of angles used when $U(\vec{\gamma_\ell}, C)$ is applied for the $\ell^{th}$ time and $\vec{\beta_\ell} = (\beta_{\ell,v_1}, \beta_{\ell,v_2},   \ldots  , \beta_{\ell, v_n}) \in \mathbb{R}^n$ is the vector of $\beta$ angles for $U(\vec{\beta_\ell}, B)$ when applied for the $\ell^{th}$ time. Each $a$ denotes a clause in the CO problem and $v_j$ refers to a specific qubit in the problem formulation, and $B_v$ typically refers to the Pauli-x matrix acting on qubit $v$. Each problem can be translated into a graph by identifying each qubit with a vertex and each interaction between qubits with an edge between the appropriate vertices \cite{herrman2021gvs, cappart2021combinatorial, smith1999neural}. 

Typically, $C$ is identified with a combinatorial optimization problem, such as MaxCut. The objective function for the MaxCut problem is

\begin{equation*}
    \min_{x \in \{0,1\}^n} \sum_{ij \in E(G)} x_j(x_i-1) + x_i(x_j-1)= \min_{x \in \{0,1\}^n} \sum_{ij \in E(G)} 2x_ix_j - x_i - x_j.
\end{equation*}
\noindent This is encoded into $C$ as

\begin{equation*}
  C=  1/2 \sum_{ij \in E(G)} (-Z_{i}Z_j + I )
\end{equation*}

\noindent where $Z_j$ is the Pauli-z operator acting on qubit $j$ and $I$ is the $2^n \times 2^n$ identity matrix. This is a diagonal matrix, so we denote the diagonal entries $d_0, \ldots, d_{2^n-1}$. Note that all entries of $C$ are zero except for possibly the diagonal entries $d_1$ through $d_{2^n-2}$, depending on the edges in the graph.

\subsection{Dynamic CTQWs}\label{sec:dctqw}

Continuous-time quantum walks (CTQWs) on graphs were introduced to search decision trees \cite{Farhi1998} and have numerous applications including modeling coherent transport on complex networks \cite{mulken2011continuous} and spatial searches \cite{childs2004spatial}. Furthermore, they are universal for computation \cite{Childs2009}. In a CTQW on a graph $G = (V,E)$, a walker moves between vertices of $G$ according to the Schr\"odinger equation

\begin{equation*}
	\label{eq:Schrodinger}
	i \frac{d\ket{\psi}}{dt} = H \ket{\psi},
\end{equation*} 
\noindent where $\hbar =1$, and $H$ is either equal to the adjacency matrix or Laplacian of $G$. In this work, we use the adjacency matrix formulation. 

The concept of CTQWs on dynamic graphs was introduced in \cite{herrman2019continuous} and proven to be universal for computation. A dynamic CTQW is a set of continuous-time quantum walks on a \textit{dynamic graph}. A dynamic graph is a set of ordered pairs of graphs and associated propagation times, $\mathcal{G} = \{(G_i, t_i)\}$. 
A dynamic CTQW is defined to be a sequence of CTQWs on each graph $G_i$ for the associated time $t_i$ performed in order of increasing index on some initial state. Mathematically, let $\ket{\psi_0}$ be the initial state of the walker and $\{(G_1, t_1), (G_2, t_2),   \ldots  , (G_k, t_k)\}$ be the dynamic CTQW that acts upon it. Then the final state of the walker, $\ket{\psi}$, is

\begin{equation*}
    \ket{\psi} = e^{-i A_k t_k/ ||A_k||} \ldots  e^{-i A_2 t_2/ ||A_2||}e^{-i A_1 t_1/ ||A_1||} \ket{\psi_0}
\end{equation*}
\noindent where $A_i$ is the adjacency matrix of $G_i$ and $|| A_i||$ is the spectral norm of $A_i$.

When showing dynamic CTQWs were universal for computation, the authors of \cite{herrman2019continuous} defined each $G_i$ to have $2^n$ vertices, which is the number of possible states in an $n$-qubit system. The vertices of these graphs were allowed to have self-loops, which adds phase to the quantum state, however multi-edges were not permitted. A self-loop in the dynamic graph on vertex $v$ is represented in the adjacency matrix as a $1$ on the diagonal element $d_v$. In the original dynamic CTQW formulation, any vertex that did not have an edge incident to it was required to have a self-loop, however the author of \cite{wong2019isolated} relaxed this condition so that vertices can exist in isolation without accumulating phase. We will use this relaxation throughout this paper.

We now recall the dynamic CTQWs that give the universal gate set of $H$, $T$, and $CX$ as defined in \cite{wong2019isolated}.

\subsubsection{Dynamic CTQW implementation of the $H$ gate}
\begin{figure}
    \centering
    \includegraphics{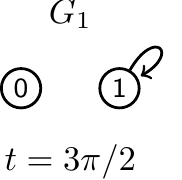} \hspace{1em}
    \includegraphics{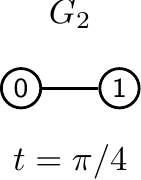} \hspace{1em}
    \includegraphics{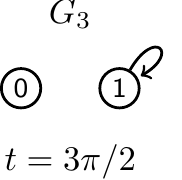}
    \caption{The dynamic CTQW implementation of the $H$ gate, which requires three graphs, $G_1$, $G_2$, and $G_3$.}
    \label{fig:Hgate}
\end{figure}

The dynamic CTQW equivalent of the single qubit $H$ gate consists of three graphs \cite{wong2019isolated} which are shown in Figure~\ref{fig:Hgate}. Formally, the dynamic CTQW for this gate is written as  $\mathcal{G}_H = \{(G_1, 3\pi/2),(G_2, \pi/4),(G_3, 3\pi/2)\}$, and the adjacency matrices corresponding to $G_1$, $G_2$, and $G_3$ are $A_1$, $A_2$, and $A_3$, respectively. These adjacency matrices are

\begin{equation*}\label{eq:selfloops}
\centering
A_1=A_3=
\begin{pmatrix}
0 & 0 \\
0 & 1  
\end{pmatrix}
\hspace{1em}
A_2=
\begin{pmatrix}
0 & 1 \\
1 & 0  
\end{pmatrix}.
\end{equation*}

\subsubsection{Dynamic CTQW implementation of the $T$ gate}
\begin{figure}
    \centering
    \includegraphics{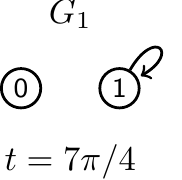}
    \caption{The dynamic CTQW implementation of the $T$ gate.}
    \label{fig:Tgate}
\end{figure}

The dynamic CTQW equivalent of the $T$ gate consists of a single graph \cite{wong2019isolated}, which is shown in Figure~\ref{fig:Tgate}. The dynamic CTQW for this gate is written as  $\mathcal{G}_T = \{(G_1, 7\pi/4)\}$, and the adjacency matrix corresponding to $G_1$ is

\begin{equation*}\label{eq:selfloops}
\centering
A_1=
\begin{pmatrix}
0 & 0 \\
0 & 1  
\end{pmatrix}.
\end{equation*}

\subsubsection{Dynamic CTQW implementation of the $CX$ gate}
\begin{figure}
    \centering
    \includegraphics{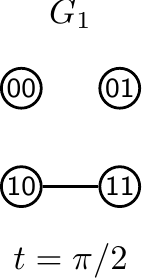} \hspace{1em}
    \includegraphics{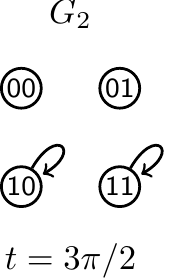} 
    \caption{The dynamic CTQW implementation of the $CX$ gate, which requires two graphs, $G_1$ and $G_2$.}
    \label{fig:CXgate}
\end{figure}

The dynamic CTQW equivalent of the two qubit $CX$ gate consists of two graphs \cite{wong2019isolated} which are shown in Figure~\ref{fig:CXgate}. Formally, the dynamic CTQW for this gate is written as  $\mathcal{G}_{CX} = \{(G_1, \pi/2),(G_2, 3\pi/2)\}$, and the adjacency matrices corresponding to $G_1$ and $G_2$ are $A_1$ and $A_2$, respectively. These adjacency matrices are

\begin{equation*}\label{eq:selfloops}
\centering
A_1=
\begin{pmatrix}
0 & 0 & 0  & 0 \\
0 & 0 & 0 & 0  \\
0 & 0 & 0 & 1  \\
0 & 0 & 1  & 0
\end{pmatrix}
\hspace{1em}
A_2=
\begin{pmatrix}
0 & 0 & 0  & 0 \\
0 & 0 & 0 & 0  \\
0 & 0 & 1 & 0  \\
0 & 0 & 0  & 1
\end{pmatrix}.
\end{equation*}

\subsection{ma-QAOA as a restriction of dynamic CTQWs}\label{sec:equivalence}

Both ma-QAOA and dynamic CTQWs act on an initial state $\ket{\psi}$ by applying operators of the form $e^{-i A_j t_j}$ where $A_j$ is a Hermitian matrix and $t_j$ is a real valued parameter. Both methods use underlying graphs to determine the structure of $A_j$, as well. Furthermore, all dynamic CTQWs for the gate set $H$, $T$, and $CX$ consist of alternating a graph that has self-loops as its only edges with graphs that have no self-loops as edges. Multi-angle QAOA is similar in that the $C$ matrix contains only diagonal entries, which is the adjacency matrix of a graph with only self-loops, up to constants, while the $B$ matrix has only off-diagonal positions. The main difference is that in ma-QAOA, the two operators $B$ and $C$ have well-defined structure that does not change with the number of iterations of the algorithm. In contrast, the dynamic CTQW graphs have no such restriction. 

Thus, it is natural to think of ma-QAOA as a restriction of dynamic CTQWs where alternate graphs have adjacency matrix $C$ and the others have adjacency matrix $B$. In order to explicitly relate ma-QAOA to dynamic CTQWs, one can define the ma-QAOA $C$ matrix as a sum of matrices, $\sum_a C_a$, each of which receives its own angle $\gamma_a$. One can also define a $B = \sum_d B_d$ matrix as a sum of matrices that receives its own angle $\beta_d$. If $\sum C_a \gamma_a = A_k t_k$, then $e^{-i \sum_a C_a \gamma_a }= e^{-i A_k t_k } $, so $U(\gamma, C)$ acts on a quantum state the same way that $ e^{-i A_k t_k } $ does, which clearly holds when considering $\sum_d B_d$ and $\beta_d$, as well. Thus, we can show that ma-QAOA is universal for computation if we can develop $B$ and $C$ matrices and find appropriate $\vec{\beta}$ and $\vec{\gamma}$ such that $\sum B_d \beta_d = A_k t_k$ or $\sum C_a \gamma_a = A_k t_k$ for each graph $A_k$ and associated time $t_k$ in the dynamic CTQW representation of the $H$, $T$, and $CX$ gates.

When implementing the dynamic CTQWs, there are two types of graphs- graphs in which a subset of vertices has self-loops with no other edges, and graphs that contain only edges with no self-loops. When solving the MaxCut problem with ma-QAOA, $C$ is a diagonal matrix with non-negative entries, which can be seen as the adjacency matrix of a graph with only weighted self-loops. $B$ has only off-diagonal entries that are 0-1 valued, which can be seen as the adjacency matrix of a graph with no self-loops. 

Since $C$ is a diagonal matrix, we want to relate $C$ to the dynamic CTQW matrices that have only self-loops as edges. The first and last adjacency matrices in the dynamic CTQW implementation of the $H$ gate, the only matrix in the dynamic CTQW implementation of the $T$ gate, and the last matrix in the dynamic CTQW implementation of the $CX$ gate are the only matrices that satisfy these criteria. Note that the self-loops for these graphs do not necessarily appear on the same vertices, so when defining $C$, we require that an arbitrary diagonal position of $C$ is the only non-zero entry of $C$. Thus, we want to define $C$ such that there exists a collection of angles $\gamma_a$ associated with each $C_a$ such that $e^{-i \sum_a C_a \gamma_a} = e^{-i A_k t_k}$ where $A_k$ is an arbitrary diagonal matrix.

Similarly, one can define a $B$ matrix as a sum of matrices, $\sum_d B_d$, each of which receives its own angle, $\beta_d$. We want to relate $B$ to the dynamic CTQW graphs that do not contain self-loops, which are the second graph in the $H$ gate implementation and the first graph in the $CX$ gate implementation. Thus, we want to define $B$ such that there exists a collection of angles $\beta_d$ associated with each $B_d$ such that $e^{-i \sum_d B_d \beta_d} = e^{-i A_k t_k}$ where $A_k$ is a matrix of the form used in the dynamic CTQW implementation of the $H$ and $CX$ gates. An advantage of relating ma-QAOA to dynamic CTQWs is that well-studied CTQW phenomena, such as hitting times, could potentially be used to understand ma-QAOA better and could potentially be used to find optimal ma-QAOA $\beta$ parameters.

\section{Using Ma-QAOA to implement a universal gate set}\label{sec:universalset}


In this section, we show how the ma-QAOA operators $B$ and $C$ and parameters $\vec{\beta}$ and $\vec{\gamma}$ can be selected that implement the equivalent of the dynamic CTQW universal set of gates $H$, $T$, and $CX$. 

\subsection{Defining $C$}

 First, we discuss how to define the $C$ operator. In the dynamic CTQW formulation, phase can be added to arbitrary vertices in the graph. 
 The MaxCut $C$ formulation in Section~\ref{sec:maqaoa}, however, does not allow for this; for example, since the diagonal entries corresponding to $\ket{0\ldots 0}$ and $\ket{1\ldots1}$ are always $0$, no phase can ever be added to these vertices. Thus, the MaxCut definition of $C$ from Section~\ref{sec:maqaoa} is not sufficient to implement the universal gate set. 
In order to ensure that phase can be added to any vertex, we modify $C$ to obtain 


\begin{equation*}
  C=  \sum_z \ket{z}\bra{z}.
\end{equation*}

\noindent In terms of the operators from Section~\ref{sec:maqaoa}, each clause $C_a$ is defined as $C_a=\ket{a}\bra{a}$. The following property of $C$ is needed to implement the universal gate set $H$, $T$, and $CX$.
\begin{property}\label{prop1}
Let $z$ be a base-10 number and $z_0  \ldots z_{n-1}$ its binary representation. There exists a linear combination of summands of $C$ with coefficients in $\{ 0, 1\}$ and 
the only non-zero diagonal entry of the linear combination is in position $z$.
\end{property}

\begin{proof}
By definition, $\ket{z} = \ket{z_0} \otimes \ldots \otimes \ket{z_{n-1}}$, and $\ket{z_i}\bra{z_i} = ((-1)^{z_i} Z_i + I_i)/2$ for $z_i \in \{0,1\}$, as  

\begin{equation*}
(Z_i + I_i)/2= [\ketbra{0}{0}+\ketbra{1}{1} +(\ketbra{0}{0}-\ketbra{1}{1})]/2 = \ketbra{0}{0}
\end{equation*} 
\noindent and 
\begin{equation*}
(-Z_i + I_i)/2= [\ketbra{0}{0}+\ketbra{1}{1} -(\ketbra{0}{0}-\ketbra{1}{1})]/2 = \ketbra{1}{1}.
\end{equation*} 


\noindent If vertex $z$ is the only vertex to receives phase, then $C_z$ receives a coefficient of $1$, while all other summands of $C$ receive a coefficient of $0$.  This ensures that $z$ is the only non-zero entry of the linear combination.
\end{proof}

Thus, phase can be added to an arbitrary vertex by picking the appropriate linear combination of summands of $C$. 
One potential drawback to this formulation of $C$ is that all combinations of $Z$ gates are required, which can lead to deep circuits. However, the diffusion operation in Grover's algorithm uses $Z$ gates and can be implemented in a scalable manner \cite{brickman2005implementation}, so we expect that the above $C$ can be implemented on large systems, as well. 


\subsection{Defining $B$}
We now turn our attention to defining $B$. 
The $T$ gate requires only self-loops, so it does not need to be considered when developing the $B$ matrix. The $B$ matrix requires summands such that a linear combination of the summands results in the adjacency matrix of a discrete hypercube, which is needed to implement the $H$ gate. In order to implement a $CX$ gate using $B$, a linear combination of the summands must also result in non-zero entries at positions $(a,b)$ and $(b,a)$ such that the binary representations of $a$ and $b$ have a 1 in position $j$ and the binary representations of $a$ and $b$ are identical except in position $v$. This must be possible for all choices of $j$ and $v$.

In order to begin constructing $B$, first note that the sum of all Pauli-x matrices acting on a single qubit, $\sum_i X_i$, is a matrix that is identical to the adjacency matrix of the discrete hypercube graph. Thus, we include $\sum_i X_i$ in the definition of $B$. Unfortunately, $CX$ is not easily implemented with just $X_i$. Each $X_i$ term connects vertices that contain a $0$ in position $i$ in its binary representation to a vertex that contains $1$ in position $i$ in its binary representation, where the rest of the terms in the binary representation are identical. 
For example, consider a two qubit system. Let $B = X_1 + X_2$, where $X_i$ is the Pauli-x operator acting on qubit $i$. Then 
\begin{equation*}
B=
\centering
\begin{pmatrix}
0 & 1 & 1  & 0 \\
1 & 0 & 0 & 1  \\
1 & 0 & 0 & 1  \\
0 & 1 & 1  & 0
\end{pmatrix} .
\end{equation*}

\begin{figure}
 \centering
 \begin{tikzpicture}[thick, main node/.style={draw, circle, inner sep=0.02cm, minimum size=0.4cm, font=\sffamily\footnotesize}, label node/.style={inner sep=0cm}]
\begin{scope}[every node/.style={scale=1.25,circle,draw}]
    \node (A) at (0,0) {$00$};
    \node (B) at (0,-1.5) {$01$};
	\node (C) at (1.5,0) {$10$}; 
	\node (D) at (1.5,-1.5) {$11$};

\end{scope}

\draw  (A) -- (B);
\draw  (A) -- (C);
\draw  (B) -- (D);
\draw  (C) -- (D);

\end{tikzpicture}
	\caption{The graph given by the adjacency matrix $B = X_1 + X_2$.}\label{fig:adjacencymatrixgraph}
\end{figure}
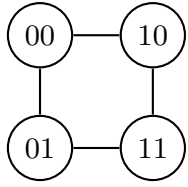

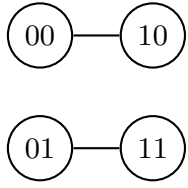
\begin{figure}
 \centering
 \begin{tikzpicture}[thick, main node/.style={draw, circle, inner sep=0.02cm, minimum size=0.4cm, font=\sffamily\footnotesize}, label node/.style={inner sep=0cm}]
\begin{scope}[every node/.style={scale=1.25,circle,draw}]
    \node (A) at (0,0) {$00$};
    \node (B) at (0,-1.5) {$01$};
	\node (C) at (1.5,0) {$10$}; 
	\node (D) at (1.5,-1.5) {$11$};

\end{scope}

\draw  (A) -- (C);
\draw  (B) -- (D);

\end{tikzpicture}
	\caption{The graph given by the adjacency matrix $B = X_1$.}\label{fig:adjmatrixX1}
\end{figure}

\noindent This is the adjacency matrix of the graph in Figure~\ref{fig:adjacencymatrixgraph}. If $B = X_1$, then  

\begin{equation*}
B=
\centering
\begin{pmatrix}
0 & 0 & 1  & 0 \\
0 & 0 & 0 & 1  \\
1 & 0 & 0 & 0  \\
0 & 1 & 0  & 0
\end{pmatrix},
\end{equation*}
\noindent which is the adjacency matrix for the graph in Figure~\ref{fig:adjmatrixX1}. In fact, if $B$ has a summand $X_i$, then it is the adjacency matrix for a graph in which the vertex corresponding to $\ket{0\ldots0}$ is adjacent to some other vertex in the graph. However, $\ket{0\ldots0}$ will never be affected by a $CX$ gate. Thus, we require a method that removes any edge incident to $\ket{0 \ldots 0}$, and potentially other edges that are not used in the $CX$ gate. 

One method of eliminating these edges is to subtract terms of the form $X_iZ_j$ to $X_i$. For example, if $B = X_1 - X_1Z_2$, then 

\begin{equation*}
B=
\centering
\begin{pmatrix}
0 & 0 & 0  & 0 \\
0 & 0 & 0 & 2  \\
0 & 0 & 0 & 0  \\
0 & 2 & 0  & 0
\end{pmatrix} .
\end{equation*}
 \noindent When divided by two, this gives the adjacency matrix of the graph in Figure~\ref{fig:adjmatrixX1Z2}, which connects vertices that are swapped in a $CX$ with control bit 2 and target bit 1.
 
 \begin{figure}
 \centering
 \begin{tikzpicture}[thick, main node/.style={draw, circle, inner sep=0.02cm, minimum size=0.4cm, font=\sffamily\footnotesize}, label node/.style={inner sep=0cm}]
\begin{scope}[every node/.style={scale=1.25,circle,draw}]
    \node (A) at (0,0) {$00$};
    \node (B) at (0,-1.5) {$01$};
	\node (C) at (1.5,0) {$10$}; 
	\node (D) at (1.5,-1.5) {$11$};

\end{scope}

\draw  (B) -- (D);

\end{tikzpicture}
	\caption{The graph given by the adjacency matrix $B = X_1 - X_1Z_2$.}\label{fig:adjmatrixX1Z2}
\end{figure}

We will now show that $B$ of the form 
\begin{equation*}
B = 1/2 (\sum_v X_v - \sum_v \sum_{j \neq v} X_v Z_j),
\end{equation*}
\noindent has the following property that will be needed to implement the $CX$ gate. 

\begin{property}\label{prop2}
There exists a linear combination of summands of $B$, called $B'$, with coefficients in $\{0,1\}$ such that for arbitrary qubits $1 \leq v \leq n$ and $1 \leq j \leq n$ where $v\neq j$, $B' = \ketbra{0_v1_j}{1_v1_j} +\ketbra{1_v1_j}{0_v1_j}$.
\end{property}

\begin{proof}
Let us show that $1/2(X_j - X_jZ_v)$ yields a matrix with the above property. 


\begin{align*}
  (  X_v - X_vZ_j) &= (\ketbra{0_v}{1_v} + \ketbra{1_v}{0_v}) - [(\ketbra{0_v}{1_v} + \ketbra{1_v}{0_v}) \otimes (\ketbra{0_j}{0_j}-\ketbra{1_j}{1_j})] \\
    &= (\ketbra{0_v}{1_v} + \ketbra{1_v}{0_v}) \otimes (\ketbra{0_j}{0_j}+\ketbra{1_j}{1_j} - (\ketbra{0_j}{0_j} - \ketbra{1_j}{1_j})) \\
    &= (\ketbra{0_v}{1_v} + \ketbra{1_v}{0_v}) \otimes 2\ketbra{1_j}{1_j} \\
    &=2 (\ketbra{0_v1_j}{1_v1_j} + \ketbra{1_v1_j}{0_v1_j}).
\end{align*}
\noindent Dividing by two gives the result.
\end{proof}






\subsection{Selecting $\vec{\beta}$ and $\vec{\gamma}$ for the $H$ gate}
The Hadamard gate, $H$, acts on a single qubit and can be represented by the matrix
\begin{equation*}
H = \frac{1}{\sqrt{2}} \left(
\begin{array}{cc}
1 & 1 \\
1 & -1 
\end{array}
\right).
\end{equation*}

\noindent The authors of \cite{herrman2022simplifying} showed that the $H^{\otimes n}$ gate can be implemented in the dynamic CTQW framework using a sequence consisting of self-loops, an $n$-dimensional hypercube, and more self-loops. The self loops can be implemented by the $C$ operators described earlier in this section and are used to add phase $\omega_v$ to vertex $v$ based on its Hamming distance from $\ket{0 \ldots 0}$. Let $h(v)$ be the Hamming distance of vertex $v$ relative to $\ket{ 0 \ldots 0}$. The phase factor required to implement the $H$ gate is

	\[ \omega_v = \begin{cases}
		-1, & h(v) \cong 0 \pmod 4 \\
		-i, & h(v) \cong 1 \pmod 4 \\
		1, & h(v) \cong 2 \pmod 4 \\
		i, & h(v) \cong 3 \pmod 4 \\
	\end{cases}\]

\noindent which is derived in \cite{herrman2022simplifying}. Since $||C_i|| =1$ for all $i$, the angles required to obtain the above phases are $\vec{\gamma_{1}} = \vec{\eta} = (\eta_1, \ldots)$, where $\eta_v$ is

	\[ \eta_v = \begin{cases}
		\pi, & h(v) \cong 0 \pmod 4 \\
		\pi/2, & h(v) \cong 1 \pmod 4 \\
		0, & h(v) \cong 2 \pmod 4 \\
		3\pi/2, & h(v) \cong 3 \pmod 4. \\
	\end{cases}\]

\noindent The $B$ operator defined earlier in this section can be used to implement the $n$-dimensional hypercube. For the $n$-dimensional hypercube, we set all $\vec{\beta_1} = (n\pi/4, \ldots, n\pi/4, 0, \ldots, 0)$, where all $n\pi/4$ angles correspond to $X_i$ terms in the sum and all $0$ angles correspond to terms of the form $X_iZ_j$. Note that $n\pi/4$ is the dynamic CTQW hypercube mixing time for the $H$ gate. For the last self-loop graph, we let $\vec{\gamma_{2}} = \vec{\gamma_{1}}$.

\subsection{Selecting $\vec{\beta}$ and $\vec{\gamma}$ for the $T$ gate}
The $T$ gate acts on a single qubit and is represented by the matrix
\begin{equation*}
T = \left(
\begin{array}{cc}
1 & 0 \\
0 & e^{\frac{i\pi}{4}} 
\end{array}
\right)
\end{equation*}

\noindent The $T$ gate acting on qubit $k$ adds a phase to all vertices that have a $1$ in position $k$ in their binary representation. This can be implemented by finding the linear combinations of summands of $C$ that give non-zero $d_i$ where $d_i$ has a $1$ in position $k$ of its binary representation and setting  $\gamma_i = 7\pi/4$, since $-7\pi/4 = \pi/4$, while setting the rest of the $\gamma_j$ to $0$. Note that in order to implement this single qubit gate with ma-QAOA, we use $C_a$, which depends on multi-qubit interactions. This is because $C$ must be able to add self-loops to arbitrary vertices in order to implement the $H$ gate, and $C$ is not allowed to change from iteration to iteration of ma-QAOA.


\subsection{Selecting $\vec{\beta}$ and $\vec{\gamma}$ for the $CX$ gate}
The controlled-not gate, denoted $CX$, is a two qubit gate. The matrix representation when the control is the first qubit and the target is the second qubit is
\begin{equation*}
\textrm{CX} =\left(
\begin{array}{cccc}
1 & 0 & 0 & 0 \\
0 & 1 & 0 & 0 \\
0 & 0 & 0 & 1 \\
0 & 0 & 1 & 0 
\end{array}
\right).
\end{equation*}

\noindent The dynamic CTQW for this gate is easy to implement and requires one graph that adds phase to a subset of vertices and another which connects vertices based on which bit is the control and which is the target. These two graphs can be performed in any order. 
Property~\ref{prop1} and Property~\ref{prop2} can be used to implement the $CX$ gate with ma-QAOA, and the implementation is similar to the dynamic CTQW one. From Property~\ref{prop2}, $e^{- i (X_j-X_jZ_v) \pi/2}$ gives the adjacency matrix for the $CX$ gate with control bit $v$ and target bit $j$, up to a factor of $-i$ on the qubits affected by the swap. The $-i$ factor is eliminated by adding a phase of $i$ to the vertices affected by the swap via Property~\ref{prop1}.

\section{Example}\label{sec:example}

\begin{figure}
\begin{center}
	\includegraphics[scale=1.5]{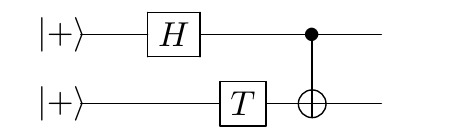} 

\end{center}
\caption{The example circuit.}
\label{fig:examplecircuit}
\end{figure}

Let us examine the circuit in Figure~\ref{fig:examplecircuit}. It has two qubits and an $H$ gate acts on the first qubit, a $T$ on the second, and a $CX$ targets the second qubit and is controlled by the first. This will require $4$ layers of ma-QAOA. The angles that implement this circuit with ma-QAOA are found in Table~\ref{tab:maqaoaanglevalues}.

\begin{figure*}[t]
\begin{center}
	\label{fig:fullcircuit}
			\includegraphics[scale=1.25]{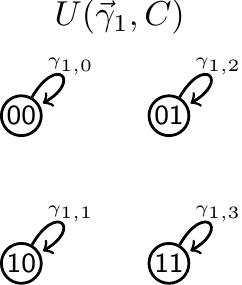} \thinspace
			\includegraphics[scale=1.25]{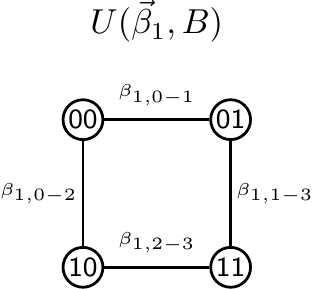} \thinspace
			\includegraphics[scale=1.25]{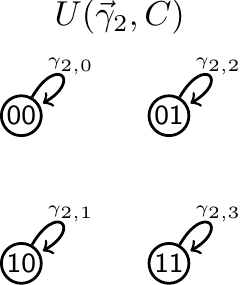} \thinspace
			\includegraphics[scale=1.25]{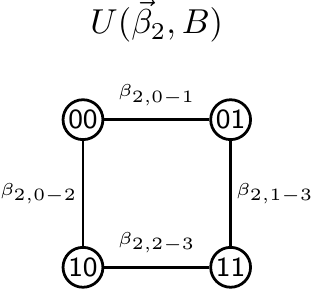}  \hfill
			
			\vspace{2em}
			\includegraphics[scale=1.25]{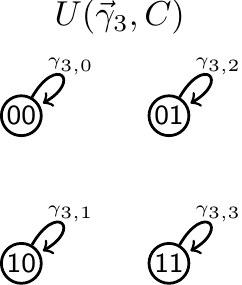} \thinspace
			\includegraphics[scale=1.25]{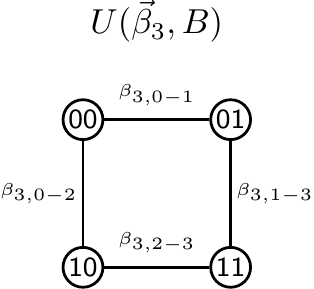} \thinspace
			\includegraphics[scale=1.25]{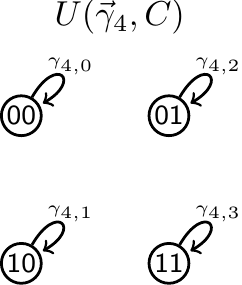} \thinspace
			\includegraphics[scale=1.25]{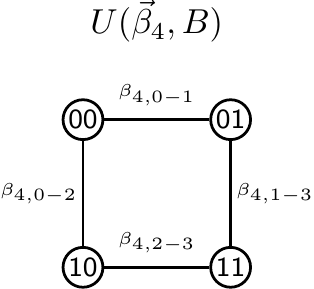} 

	\caption{The ma-QAOA implementation of the circuit found in Figure~\ref{fig:examplecircuit} with angles found in Table~\ref{tab:maqaoaanglevalues}. The top row implements the $H$ gate, the left graph on the bottom row implements the $T$ gate, and the bottom middle two graphs implement the $CX$ gate. The last graph is not necessary, but is included to show four full iterations of ma-QAOA.}
\end{center}
\end{figure*}

\begin{table}\footnotesize
    \centering
    \begin{tabular}{|c|c|}
        \hline
       $U$  & Angle vector    \\
       
        \hline
          $U(\gamma_1, C)$ & $(\pi,\pi,  \pi/2, \pi/2)$   \\
        \hline
         $U(\beta_1, B)$ &   $(0, \pi/4, \pi/4, 0)$ \\
        \hline
         $U(\gamma_2, C)$ &  $(\pi,\pi,  \pi/2, \pi/2)$    \\
        \hline
         $U(\beta_2, B)$ &  $(0, 0, 0, 0)$ \\
        \hline
         $U(\gamma_3, C)$ &  $(0, 7\pi/4, 0, 7\pi/4)$  \\
        \hline
         $U(\beta_3, B)$ &  $(0, 0, 0, 3\pi/2)$  \\
        \hline 
        $U(\gamma_4, C)$ &  $(0,0,  \pi/2, \pi/2)$  \\
        \hline
         $U(\beta_4, B)$ & $(0, 0, 0, 0)$  \\
        \hline

    \end{tabular}
    \caption{The $\beta_{p,i-j}$ angles for an ma-QAOA implementation of the circuit found in Figure~\ref{fig:examplecircuit}. The vectors $\gamma_i$ have entries $(\gamma_{i,0},\gamma_{i,1},\gamma_{i,2},\gamma_{i,3})$ and vectors $\beta_j$ have entries $(\beta_{j,0-1},\beta_{j,0-2},\beta_{j,1-3},\beta_{j,2-3})$, where the subscript $f-g$ for $f, g \in \mathbb{R}$ refers to the edge connecting vertex $f$ to vertex $g$.
    }
    \label{tab:maqaoaanglevalues}
\end{table}

We now confirm that these angles implement the above circuit when used with the $B$ and $C$ operators as defined above. First, note that $H \otimes I$ followed by $I \otimes T$ followed by $CX$ with control qubit $1$ and target qubit $2$ acts on the state $\ket{+}$ as
	\begin{align*}
		& 1/2\big[ \ket{00} +  \ket{01} +  \ket{10} + \ket{11} \big]\\
		&\quad \xrightarrow{H \otimes I} 1/\sqrt{2}\big[ \ket{00} + \ket{01} \big] + 0 \big[ \ket{10} + \ket{11}\big] \\
		&\quad \xrightarrow{I \otimes T}  1/\sqrt{2}\big[ \ket{00} + e^{i \pi/4}\ket{01} \big] + 0 \big[ \ket{10} + e^{i \pi/4}\ket{11}\big] \\
		&\quad \xrightarrow{CX}  1/\sqrt{2}\big[ \ket{00} + e^{i \pi/4}\ket{11} \big] + 0 \big[ e^{i \pi/4}\ket{01} + \ket{10}\big] \\ 
	\end{align*}

	The ma-QAOA implementation of the $H$ gate varies slightly from above, as $H$ only acts on one qubit in this circuit, not both of them. The 1-dimensional hypercube is a path on two vertices, and since the $H$ is performed on the first qubit, the non-zero $\beta$ angles correspond to edges that connect vertices whose first qubit are either both $0$ or both $1$. Additionally, the $\omega_v$ are changed so that one vertex in each two-dimensional hypercube has a factor of $-1$ and the other has a factor of $-i$. The angles for this implementation are found in Table~\ref{tab:maqaoaanglevalues}. These angles act on the initial state $\ket{+}$ as
	
	\begin{align*}
		& 1/2\big[ \ket{00} +  \ket{01} +  \ket{10} + \ket{11} \big]\\
		&\quad \xrightarrow{U(\vec{\gamma_1}, C)} -1/2\big[\ket{00} + \ket{01}\big] - i/2\big[\ket{10} + \ket{11}\big] \\
		&\quad \xrightarrow{U(\vec{\beta_1}, B)} -1/\sqrt{2}\big[ \ket{00} + \ket{01} \big] + 0 \big[ \ket{10} + \ket{11}\big] \\
		&\quad \xrightarrow{U(\vec{\gamma_2}, C)}  1/\sqrt{2}\big[ \ket{00} + \ket{01} \big] + 0 \big[ \ket{10} + \ket{11}\big] \\
		&\quad \xrightarrow{U(\vec{\gamma_2}, B)}  1/\sqrt{2}\big[ \ket{00} + \ket{01} \big] + 0 \big[ \ket{10} + \ket{11}\big] \\
		&\quad \xrightarrow{U(\vec{\gamma_3}, C)}  1/\sqrt{2}\big[ \ket{00} +  e^{i \pi/4}\ket{01} \big] + 0 \big[ \ket{10} + e^{i \pi/4}\ket{11}\big]\\
		&\quad \xrightarrow{U(\vec{\beta_3}, B)}  1/\sqrt{2}\big[ \ket{00} -i \ket{11} \big] + 0 \big[ e^{i \pi/4}\ket{01} + e^{i \pi/4}\ket{11}\big]\\
		&\quad \xrightarrow{U(\vec{\gamma_4}, C)}  1/\sqrt{2}\big[ \ket{00} +  e^{i \pi/4}\ket{11} \big] + 0 \big[ e^{i \pi/4}\ket{01} + \ket{10}\big]\\
		&\quad \xrightarrow{U(\vec{\beta_4}, B)}  1/\sqrt{2}\big[ \ket{00} +  e^{i \pi/4}\ket{11} \big] + 0 \big[ e^{i \pi/4}\ket{01} + \ket{10}\big],
	\end{align*}
	which is the same final state as before.

\section{Discussion}\label{sec:discussion}
In this paper, we show that ma-QAOA is equivalent to a restriction of dynamic CTQWs in which the underlying graphs can only consist of singletons or of discrete hypercubes with dimension at most $n$ and then find the appropriate operators and angles that yield the universal gate set $H$, $T$, and $CX$ in the ma-QAOA framework. Since ma-QAOA can be viewed as a restriction of dynamic CTQWs, there is potential that well-studied CTQW phenomena, such as hitting times, can be used to determine the optimal $\beta$ parameters for ma-QAOA. Finding optimal QAOA parameters is a challenging problem, and several techniques such as transferability, reinforcement learning, and neural networks \cite{galda2021transferability, shaydulin2022parameter, wauters2020reinforcement, verdon2019learning}. Relating ma-QAOA to CTQWs gives a new framework through which to view the algorithm. It would be of interest to determine if viewing CTQWs through a QAOA lens leads to better understanding of CTQWs. 

The authors of \cite{herrman2022simplifying} gave methods for simplifying dynamic CTQWs based on the underlying graph structure. Since ma-QAOA is equivalent to a restricted dynamic CTQW, ma-QAOA operators can potentially be rearranged and combined to reduce the circuit depth needed to implement arbitrary operations. In the formulation in this manuscript, each $H$, $CX$, and $T$ implementation require at most 1.5 layers of ma-QAOA, so a circuit that implements $N$ of these gates would require at most $1.5N$ layers of ma-QAOA. Future work includes examining if there are operations that can be used to reduce the ma-QAOA circuit depth outside of those found in \cite{herrman2022simplifying}.



Finally, in \cite{lloyd2018quantum}, the author proves the universality of QAOA using a line graph quantum architecture, which is not easily comparable to the method of showing ma-QAOA universality in this paper. The author mentions that this architecture is limited but says that the techniques used to prove QAOA universality can be expanded to higher dimensions. It would be of interest to determine if there are cases where the QAOA universality methods in \cite{lloyd2018quantum} require fewer operations to implement an arbitrary circuit than the method in this paper, and vice-versa. Additionally, fully-connected architecture and gates that act on $n$ qubits for all $n \geq 2$ are required in this implementation, whereas the line architecture in \cite{lloyd2018quantum} is much more sparse. It would be of interest to determine if there is a more natural method of using QAOA or ma-QAOA for computation on lattices such as the square grid or hexagonal lattice, which more closely model current quantum architecture \cite{lotshaw2022scaling,paler2014mapping}.

\bibliographystyle{unsrt}
\bibliography{manuscript}

\section*{Acknowledgements}
The author would like to thank Phillip Lotshaw and Travis Humble for their feedback and insight.


\end{document}